\documentclass{PoS}

\usepackage{graphicx}
\usepackage[utf8]{inputenc}
\usepackage[english]{babel}
\usepackage{amsmath,amssymb,color,amscd,subfigure}

\providecommand*{\ord}{\ensuremath{\text{O}}}

\providecommand*{\e}{\ensuremath{\text{e}}}
\providecommand*{\De}{\ensuremath{\mathcal{D}}}

\title{Reweighting twisted boundary conditions}

\ShortTitle{Reweighting twisted boundary conditions}

\author{\speaker{A. Bussone} \\
CP3-Origins \& Danish IAS, University of Southern Denmark, Campusvej 55, 5230 Odense M.\\
        E-mail: \email{bussone@cp3-origins.net}}

\author{M. Della Morte \\
CP3-Origins \& Danish IAS, University of Southern Denmark, Campusvej 55, 5230 Odense M.\\
IFIC (CSIC), Calle Catedr\'{a}tico Jos\'{e} Beltran, 2, 46980 Paterna, Valencia, Spain\\
        E-mail: \email{dellamor@cp3-origins.net}}
        
\author{M. Hansen \\
CP3-Origins \& Danish IAS, University of Southern Denmark, Campusvej 55, 5230 Odense M.\\
        E-mail: \email{hansen@cp3-origins.net}}
        
\author{C. Pica \\
CP3-Origins \& Danish IAS, University of Southern Denmark, Campusvej 55, 5230 Odense M.\\
        E-mail: \email{pica@cp3-origins.net}}

\abstract{{\it Preprint: CP3-Origins-2015-037 DNRF90, DIAS-2015-37, IFIC/15-65}\\
\\
Imposing twisted boundary conditions on the fermionic fields is a procedure extensively used when evaluating, for example, form factors on the lattice. Twisting is usually performed for one flavour and only in the valence, and this causes a breaking of unitarity.
In this work we explore the possibility of restoring unitarity through the reweighting method. We first study some properties of the approach at tree level and then 
we stochastically evaluate ratios of fermionic determinants
for different boundary conditions in order to include them in the gauge averages, avoiding in this way the expensive generation of  new configurations for each choice of the twisting angle, $\theta$.
As expected the effect of reweighting is negligible in the case of large volumes but it is important when the volumes are small  and the twisting angles are large. In particular we find a measurable effect for the plaquette and the pion correlation function in the case of $\theta=\pi/2$ in a volume $16\times 8^3$, and we observe a systematic upward shift in the pion dispersion relation.
}

\FullConference{The 33rd International Symposium on Lattice Field Theory\\
                 14 -18 July  2015\\
                 Kobe International Conference Center, Kobe, Japan}

\begin{document}

\section{Introduction}

Non-Periodic Boundary Conditions (NPBCs) for the fermions are typically employed on the lattice in order to obtain
a fine resolution of momenta in the spatial directions. {\it Twisting}~\cite{deDivitiis:2004kq} amounts to imposing
	\begin{align*}
					\psi\left(x+N_\mu\hat{\mu}\right) =
					\begin{cases}
							\e^{i\theta_\mu}\psi(x)\;, \quad \mu=1,2,3\\
							\psi(x)\;,\;\;\,\qquad \mu=0
						\end{cases} \;,
\end{align*}
where $N_\mu$ is the lattice extent in direction $\hat{\mu}$ and $\theta_j\in\left[0, 2\pi\right]$ is an angle.
Alternatively one can introduce a constant $\mathbf{U}(1)$ interaction with vanishing electric, magnetic field and electric potential but constant vector potential \cite{Bedaque:2004kc} (see also \cite{Sachrajda:2004mi} for the equivalence of the two procedures).
The above  
extra interaction is implemented by transforming the standard QCD links ($U$) in the following way
\begin{equation}
\label{eq:modified_links}
					\mathcal{U}_\mu(x)= \begin{cases}
						\e^{i\theta_\mu/N_L}U_\mu(x)\;, \quad \mu=1,2,3\\
						U_0(x)\;, \;\;\,\quad\qquad \mu=0
					\end{cases}.
\end{equation}
\\
The modification of the boundary conditions proved to be beneficial for:
\begin{itemize}
\item Form factors: one can scan values of the exchanged momenta in the scattering process with a very fine resolution, in order to determine more accurately the form factors on the lattice. An example is the semi-leptonic decay $K_{\ell 3}$ \cite{Boyle:2007qe}, used to extract the CKM element $V_{us}$.
\item Matching between  HQET and QCD: in the Schr\"odinger functional, one computes finite volume observables for different values of $\theta$ in the valence \cite{DellaMorte:2013ega}. In particular in such a setup reweighting to a unitary formulation is expected to be efficient as the volumes 
considered are rather 
small (the reweighting factors are extensive quantities).
\item Dispersion relation: 
which is used here more as a consistency check, as done also in~\cite{deDivitiis:2004kq}.
\end{itemize}

Usually one performs the twisting only in the valence, which causes breaking of unitarity. This is simply understood since the procedure yields different propagators for fermions in the valence and in the sea. One way to overcome this problem is to perform direct simulations with fermionic NPBCs for each value of $\theta$. That though, would clearly be too expensive from the computational point of view.
However, this kind of breaking 
of unitarity is expected to be a finite volume effect, in $\chi$-PT for example that is the case \cite{Sachrajda:2004mi}. This suggests that
where the effect is large, reweighting may provide a reliable way to restore unitarity.

\section{Reweighting}

Let us suppose we want to connect simulation results for a choice of 
bare parameters $a = \{\beta, m_1, m_2, \dots, m_{n_f}, \theta_\mu, \dots\}$ to a (slightly) different set $b = \{\beta', m'_1, m'_2, \dots, m'_{n_f}, \theta'_\mu, \dots\}$ of parameters. 
This can be achieved by numerically computing on the $a$-ensemble the reweighting factor $W_{a b} = P_b /P_a$, which is the ratio of the two probability distributions and it is clearly an extensive quantity, $P_a[U] = \e^{-\text{S}_{\text{G}}[\beta, U]} \prod_{i=1}^{n_f}\det\left(D[U,\theta]+m_i\right)$, where we have explicitly indicated the dependence of the Dirac operator on the twisting angle.
In this way the expectation values on the $b$-ensemble are calculated as
\begin{equation*}
\langle\mathcal{O}\rangle_b = \frac{\langle\widetilde{\mathcal{O}} W_{a b}\rangle_a}{\langle W_{a b}\rangle_a}\, ,
\end{equation*}
with $\widetilde{\mathcal{O}}$ the observable after Wick contractions, and $\langle\dots\rangle_{a}$ the expectation value over the set of bare parameters $a$.\\
By choosing to change only the boundary conditions from one bare set to the other we arrive at the following expression of the reweighting factor 
\begin{align*}
		W_\theta = \det\left(D_W[U,\theta]D_W^{-1}[U,0]\right) = \det\left(D_W[\mathcal{U},0]D_W^{-1}[U,0]\right),
\end{align*}
where $D_W$ is the massive Wilson operator. We therfore need a stochastic method to estimate (ratio of) determinants.
For this purpose we use the following integral representation of a normal matrix determinant with spectrum $\lambda(A)$ that holds if and only if the real part of 
each eigenvalue is larger than zero \cite{Finkenrath:2013soa}
\begin{equation}
\label{eq:appl_cond}
\frac{1}{\det A}  = \int \De\left[\eta\right] \exp\left(-\eta^\dagger A \eta\right)<\infty\, \Longleftrightarrow\, \mathbb{R}\text{e}\lambda\left( A\right)>0.
\end{equation}
If it is so then the stochastic estimae converges. We use as probability distribution $p(\eta)$ of the $\eta$ vectors a gaussian one, then the determinant reads
\begin{align*}
\frac{1}{\det A}  = \bigg\langle \frac{\e^{-\eta^\dagger A \eta}}{p(\eta)} \bigg\rangle_{p(\eta)} = \frac{1}{N_\eta} \sum_{k=0}^{N_\eta} \e^{-\eta_k^\dagger (A-\mathbf{1}) \eta_k} + \ord\left(\frac{1}{\sqrt{N_\eta}}\right).
\end{align*}
It should be noted that if we require, for the case of an hermitian matrix, the existence of all gaussian moments then all the eigenvalues must be strictly larger than one
\begin{align*}
\bigg\langle \frac{ \e^{-2\eta^\dagger A \eta} }{p(\eta)^2} \bigg\rangle_{p(\eta)} & = \int \De\left[\eta\right] \exp\left[-\eta^\dagger (2A-\mathbf{1}) \eta\right] < \infty\, \Longleftrightarrow\, \lambda\left( A\right)>\frac{1}{2},\\
\vdots \hspace{1.4cm}&\\
\bigg\langle \frac{ \e^{- N \eta^\dagger A \eta} }{p(\eta)^N} \bigg\rangle_{p(\eta)} & = \int \De\left[\eta\right] \exp\left[-\eta^\dagger [NA-(N-1)\mathbf{1}] \eta\right] <\infty\, \Longleftrightarrow\, \lambda\left( A\right)>\frac{N-1}{N} \underset{N\rightarrow\infty}{\longrightarrow} 1.
	\end{align*}

\subsection{Tree level study}

\begin{figure}[h!t]
\begin{center}
\subfigure[\emph{Mean of the reweighting factor at tree level for $\theta = 0.1$ as a function of $N_L$.}]{\includegraphics[scale=0.64]{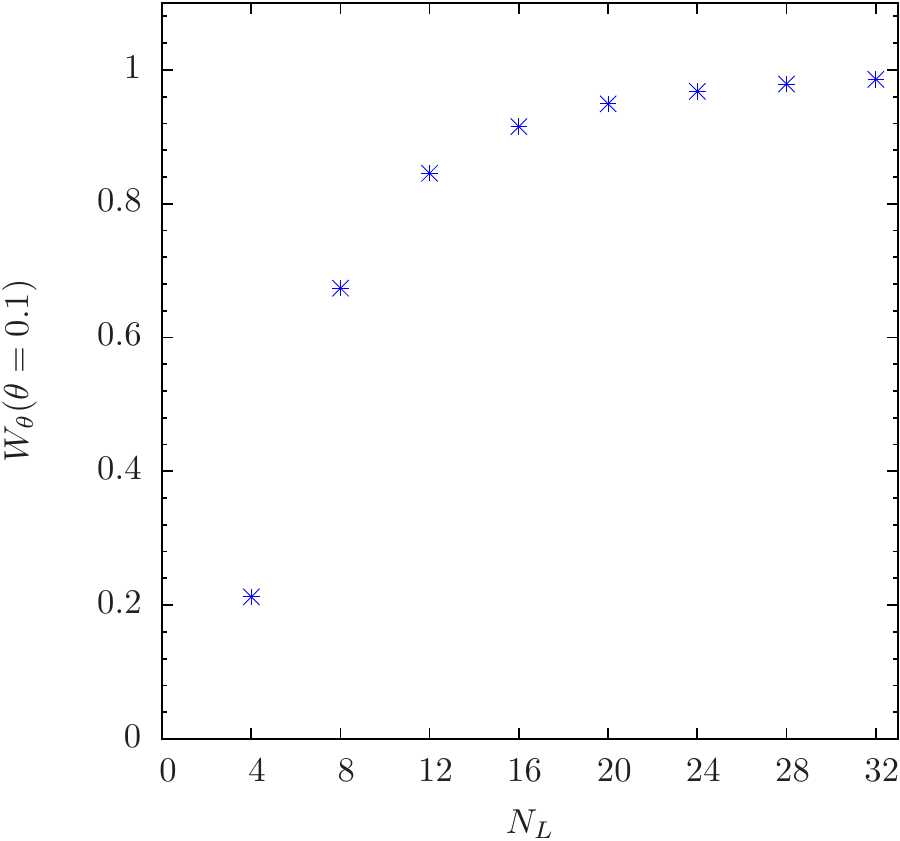}\label{fig:tree_level_test1}}
\hspace{0.5cm}
\subfigure[\emph{Variance of the reweighting factor at tree level for $\theta = 0.1$ as a function of $N_L$.}]{\includegraphics[scale=0.64]{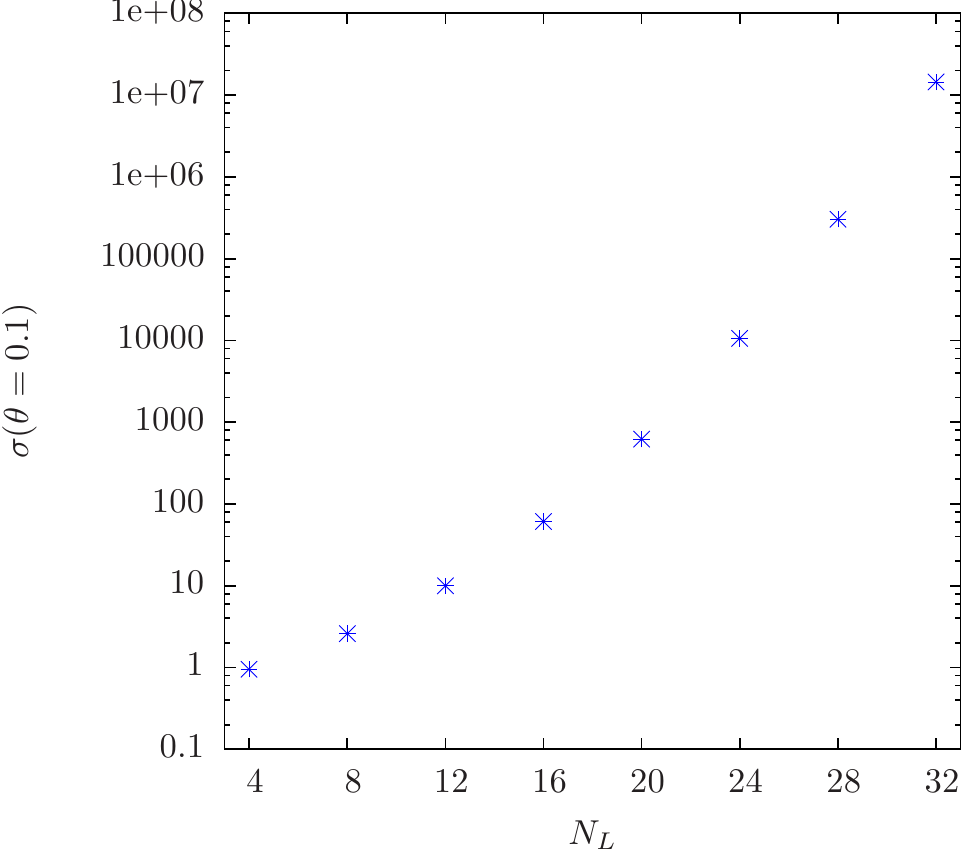}\label{fig:tree_level_test2}}
\caption{\emph{Estimates of mean and variance of the reweighting factor employing the exact formulae for the tree level case. Each point corresponds to a cubic lattice of the form $L^4$.}}
%\label{fig:tree_level_test}
\end{center}
\end{figure}

The spectrum of the Dirac-Wilson operator at tree level is known and that allowed us to test our numerical implementation. A tree level calculation
(Fig.~\ref{fig:tree_level_test1}) shows that at fixed $\theta$ and for large $N_L$ the reweighting factor approaches the value 1, which is expected since it is a finite volume effect. Conversely, the variance grows as $N_L$ increases (Fig.~\ref{fig:tree_level_test2}). Hence a direct estimate for large $\theta$ angles 
is not reliable and we need to employ a multi-step  method in order to keep the error under control.
%%%%%%%%%%%%%%%%%%%%%%%%%%%%%%%%%%%%%%%%%%%%%%%%%%%%%%%%%%%%%%%%%%%%%%%%%%
%%%%%%%%%%%%%%%%%%%%%%%%%%%%%%%%%%%%%%%%%%%%%%%%%%%%%%%%%%%%%%%%%%%%%%%%%%
To this end, following \cite{Finkenrath:2013soa}, we factorize the relevant matrix in the following telescopic way
\begin{align*}
		D_W(\theta)D_W^{-1}(0) = \prod_{l=0}^{N-1}A_\ell,\,\text{ with }A_\ell\simeq\mathbf{1}+\ord\left(\delta\theta_\ell\right)\;, \quad N=\frac{\theta}{\delta\theta_\ell}\;.
	\end{align*}
with $A_\ell$ now near to the identity matrix as it corresponds to a small
$\delta\theta_\ell$ shift in the periodicity angle. At this point the inverse determinant and its variance are given in terms of the $N$ analogous quantities,
one for each multiple of $\delta \theta_\ell$ in $\theta$, as
\begin{align*}
				\frac{1}{\det A} &= \prod_{\ell=0}^{N-1}\bigg\langle \frac{\exp\left(-\eta^{(\ell),\dagger} A_\ell \eta^{(\ell)}\right)}{p\left(\eta^{(\ell)}\right)} \bigg\rangle_{p\left(\eta^{(\ell)}\right)},\\
				\sigma^2 & = \sum_{\ell=0}^{N-1}\left[\sigma^2_{\eta^{(\ell)}}\prod_{k\neq\ell}\det \left(A_k\right)^{-2}\right].
			\end{align*}

\section{Monte Carlo studies}

\begin{figure}[h!t]
\begin{center}
\subfigure[\emph{Mean of the reweighting factor for $\theta = 0.3$ as a function of $N_\eta$.}]{\includegraphics[scale=0.7]{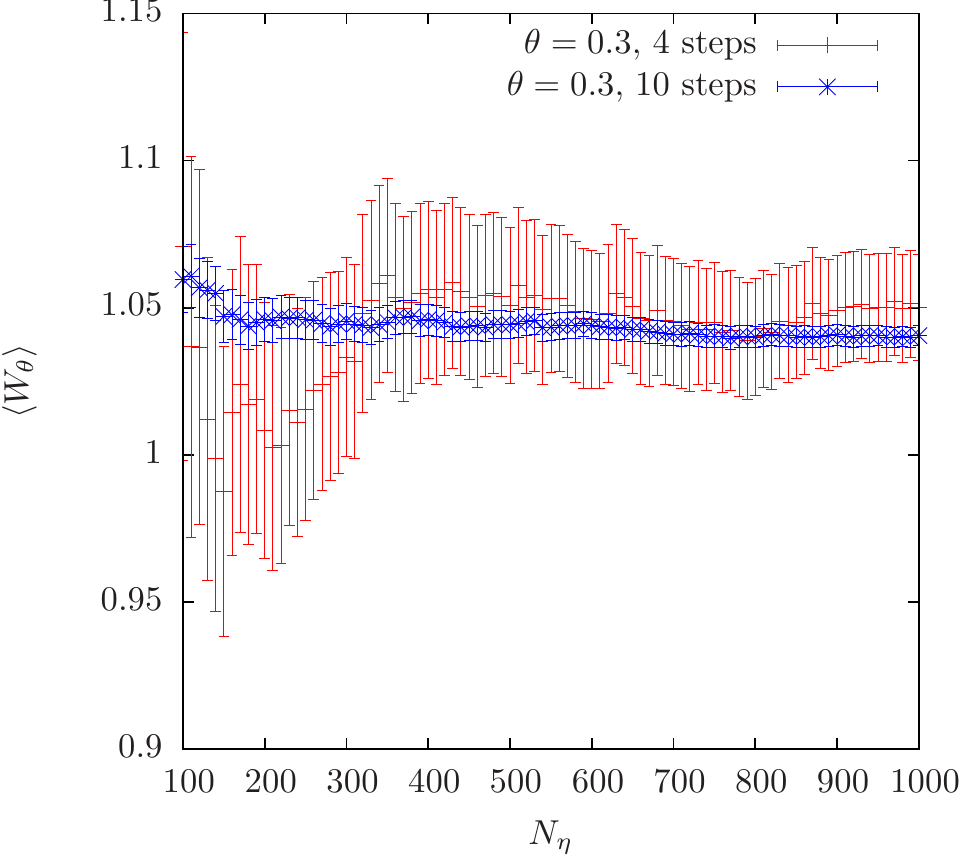}\label{fig:mc_hist_1}}
\hspace{0.5cm}
\subfigure[\emph{Mean of the reweighting factor for $\theta = \pi /2$ as a function of $N_\eta$.}]{\includegraphics[scale=0.7]{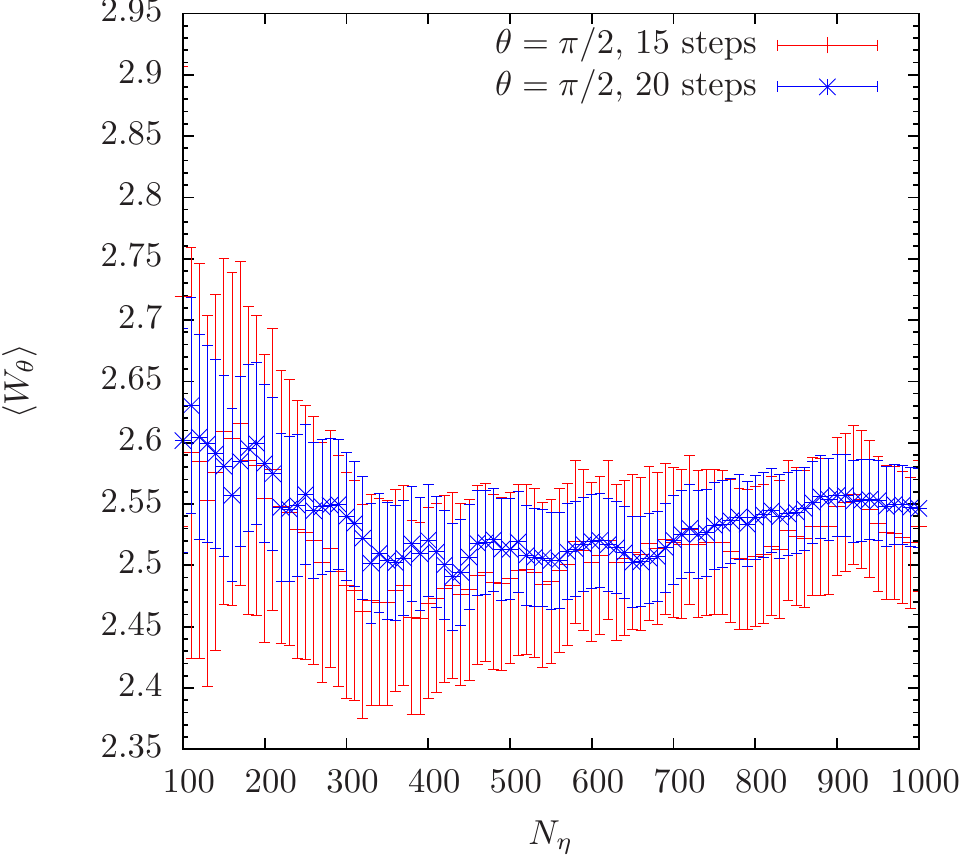}\label{fig:mc_hist_2}}
\caption{\emph{Monte Carlo history of the reweighting factor, averaged over the entire number of configurations. The figures correspond to a volume $V = 16\times 8^3$.}}

\end{center}
\end{figure}

We have calculated a number of  reweighted observables in the $\mathbf{SU}(2)$ gauge theory with fermions in the fundamental representation. This theory is known to exhibit confinement and chiral symmetry breaking and it is therefore QCD-like. We have employed unimproved Wilson fermions and the Wilson plaquette gauge action. In the table below we list the configurations used in this work ($m_\text{cr}$ is estimated to be $-0.75(1)$ at $\beta = 2.2$).
\begin{center}
	\begin{tabular}{|c|c|c|c|c|}
	\hline
	$V$ & $\beta$ & $m$ & $N_\text{cnf}$ & traj. sep.\\
	\hline
	$16 \times 8^3$ & 2.2 & -0.6 & $10^3$ & 10 \\
	\hline
	$32 \times 24^3$ & 2.2 & -0.65 & $394$ & 20 \\
	$32 \times 24^3$ & 2.2 & -0.72 & $380$ & 10 \\
	\hline
	\end{tabular}
\end{center}
We have reweighted observables to constant non-vanishing spatial $\theta$. To evaluate the reweighting factor we have used the $\gamma_5$ version of the Dirac-Wilson operator, 
since it is hermitian 
and with 2 flavours it automatically satisfies the applicability condition in eq.~\eqref{eq:appl_cond}.
We found that for $N_\eta \gtrsim 200$  the reweighting factor on each configuration is larger than 10 times the average value on a few configurations only.
This guarantees that the average values are not dominated by spikes and hence that there are no large statistical fluctuations of the determinant.
In the figures below we show the mean of the reweighting factor (all points are averaged over the total number of configurations).
In Figs.~\ref{fig:mc_hist_1} and~\ref{fig:mc_hist_2} 
a good scaling for the error is clearly visible, according to $N_\eta^{-1/2}$ and up to $N_\eta \approx 500$, when increasing the number of steps.
We conclude that at these parameters taking 500$-$600 gaussian vectors is enough to saturate the statistical noise at the level of the gauge noise.

%%%%%%%%%%%%%%%%%%%%%%%%%%%%%%%%%%%%%%%%%%%%%%%%%%%%%%%%%%%%%%%%%%%%%%%%%%
%%%%%%%%%%%%%%%%%%%%%%%%%%%%%%%%%%%%%%%%%%%%%%%%%%%%%%%%%%%%%%%%%%%%%%%%%%

	\subsection{Small volume}
	
\begin{figure}[h!t]
\begin{center}
\subfigure[\emph{Plaquette with only one flavor reweighted.}]{\includegraphics[scale=0.7]{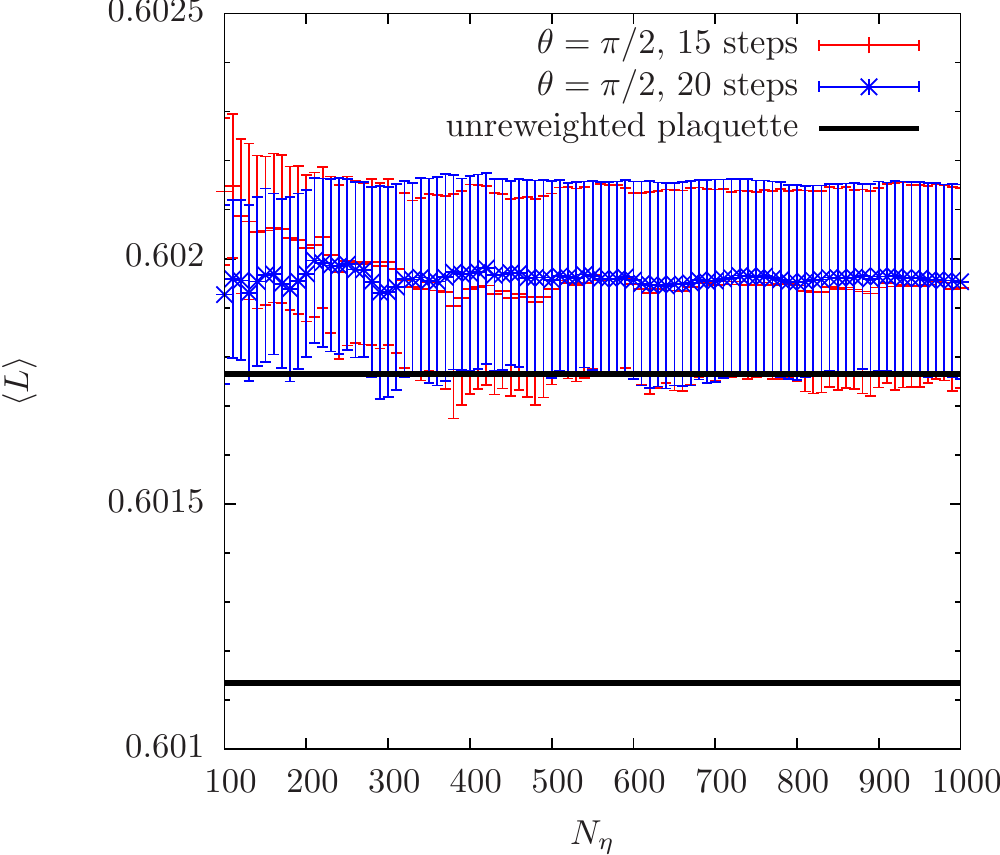}\label{fig:plaquette_small_1}}
\hspace{0.5cm}
\subfigure[\emph{Plaquette with both flavors reweighted.}]{\includegraphics[scale=0.7]{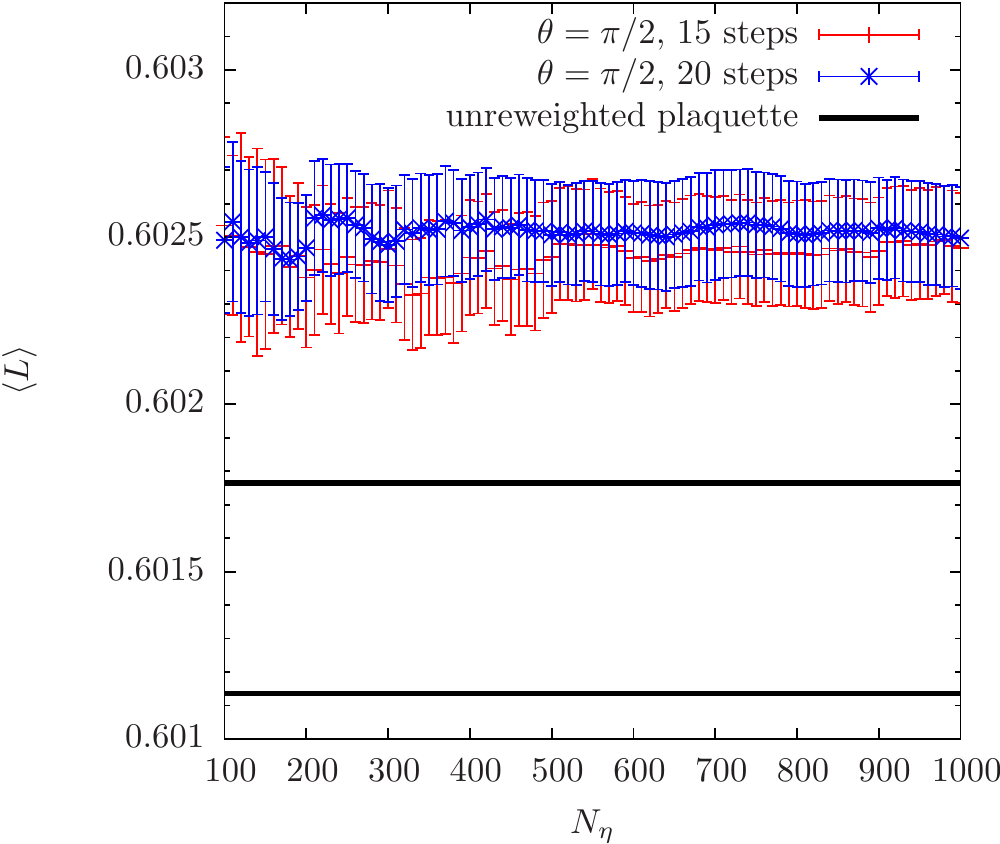}\label{fig:plaquette_small_2}}
\caption{\emph{Monte Carlo history of the reweighted plaquette. The figures correspond to a volume $V=16\times 8^3$.}}
\end{center}
\end{figure}

Reweighting may give a significant 
effect in small volumes.
We have first looked at the plaquette because it does not explicitly depend on the BCs since, in general, in Wilson loops 
for each link entering the loop there is another one in the opposite direction and the exponential factors in eq.~\eqref{eq:modified_links})
cancel among the two.
We are neglecting autocorrelations, because measurements are separated by 10 to 20 molecular dynamics units. A binning procedure, including bins from length one up to $10$, shows no significant autocorrelation effects.
In Fig.~\ref{fig:plaquette_small_1} we show results after reweighting only one flavor, by taking the square root of the reweighting factor. No effects are visible in this case within errors.
In Fig.~\ref{fig:plaquette_small_2} instead we have included the reweighting factor for both flavors and one can see a permil 
effect, recognizable because the plaquette is determined very accurately.

%%%%%%%%%%%%%%%%%%%%%%%%%%%%%%%%%%%%%%%%%%%%%%%%%%%%%%%%%%%%%%%%%%%%%%%%%%
%%%%%%%%%%%%%%%%%%%%%%%%%%%%%%%%%%%%%%%%%%%%%%%%%%%%%%%%%%%%%%%%%%%%%%%%%%

The second quantity that we have studied is the pion dispersion relation.
In the pion correlator we twisted only one flavor in the valence, hence the lowest energy state becomes a pion with momentum $\vec{p} = \pm\vec{\theta} / L$. We have reweighted the correlators 
in order to take into account the twist in the sea and we have extracted the effective mass after symmetrizing the correlators in time.

\begin{figure}[h!t]
\begin{center}
\includegraphics[scale=0.7]{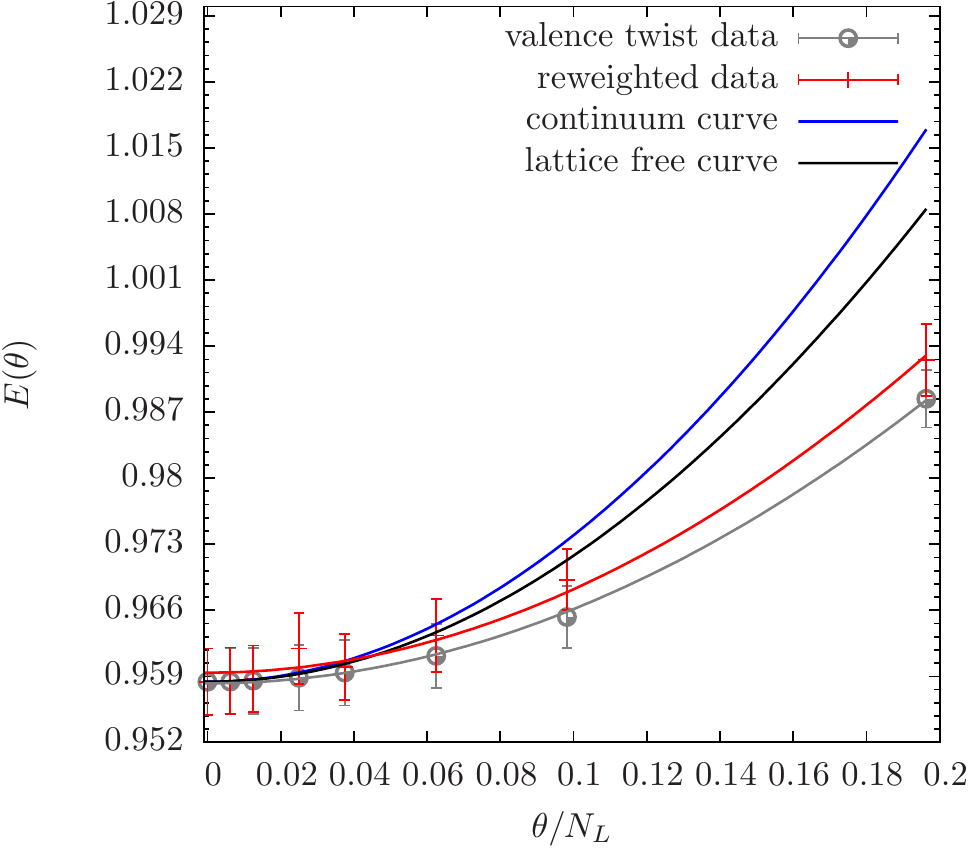}
\caption{\emph{Pion dispersion relation for $V = 16\times 8^3$. Each point corresponds to the best determination of the reweighting factor.}}
\label{fig:disp_rel_small}
\end{center}
\end{figure}

In Fig.~\ref{fig:disp_rel_small} we show results for the dispersion relation and we compare them to un-reweighted data (i.e., with twisting only in the valence), to the continuum prediction $(a E)^2 = (a m_\pi)^2 + 3 \theta^2 / N_L^2$ and to the lattice free (boson) theory prediction $\cosh(a E) = 3 + \cosh(a m_\pi) + 3 \cos(\theta / N_L)$.
Over the entire range of  $\theta$ values explored there is no clear effect within errors. However there is a systematic effect upward.
The discrepancy between the reweighted data and the lattice free prediction is conceivably due to cutoff effects and non-perturbative dynamics.

%%%%%%%%%%%%%%%%%%%%%%%%%%%%%%%%%%%%%%%%%%%%%%%%%%%%%%%%%%%%%%%%%%%%%%%%%%
%%%%%%%%%%%%%%%%%%%%%%%%%%%%%%%%%%%%%%%%%%%%%%%%%%%%%%%%%%%%%%%%%%%%%%%%%%

	\subsection{Large volumes}

\begin{figure}[h!t]
\begin{center}
\subfigure[\emph{Case of $m\simeq -0.65$ which corresponds to a heavy pion.}]{\includegraphics[scale=0.72]{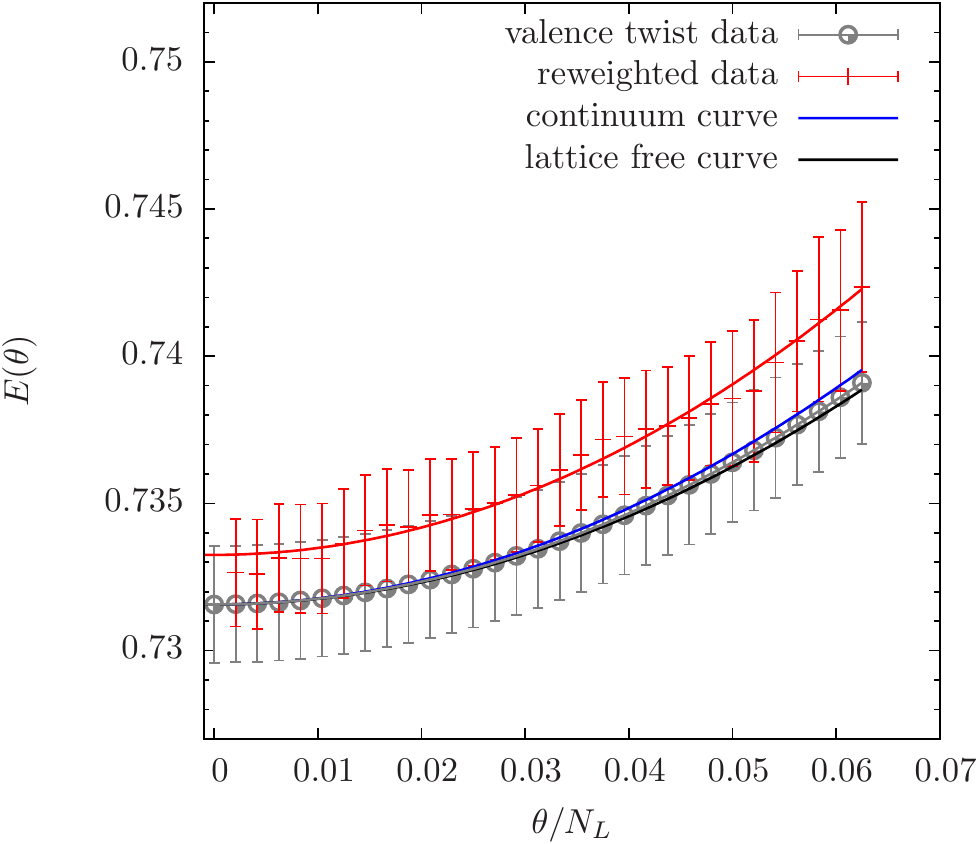}}
\hspace{0.5cm}
\subfigure[\emph{Case of $m\simeq -0.72$ which corresponds to a lighter pion.}]{\includegraphics[scale=0.72]{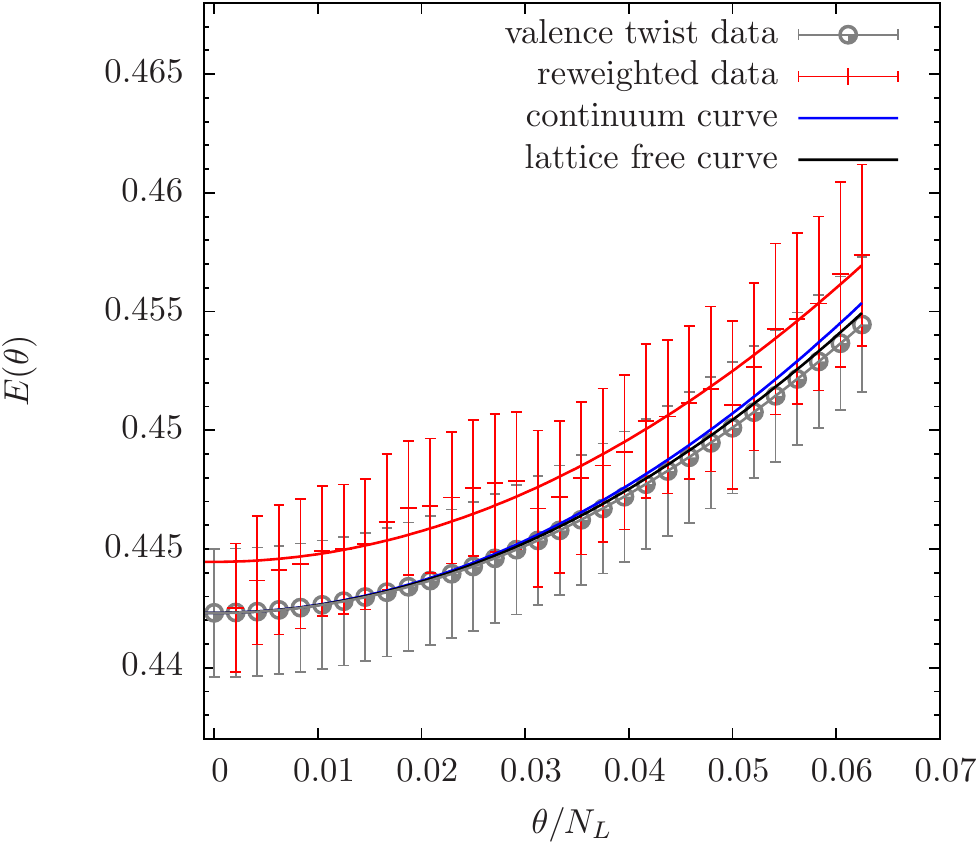}}
\caption{\emph{Pion dispersion relation for $V = 32\times 24^3$. The reweighting factor at a given $\theta$ is obtained by a telescopic product involving all the previous ones.}}
\label{fig:disp_rel_large}
\end{center}
\end{figure}

Obviously, in large volume we do not expect the situation to improve with respect to small volumes, concerning both statistical accuracy and significance of the reweighting.
We have looked at the pion dispersion relation for two rather different values of $m_\pi$. 
We expect to find a larger effect for the lighter pion but already at these volumes it appears as one can neglect the effects coming from the breaking of unitarity. 
In Fig.~\ref{fig:disp_rel_large} we show the results for the two pion masses.
No sizeable effect is detectable in either case.
%%%%%%%%%%%%%%%%%%%%%%%%%%%%%%%%%%%%%%%%%%%%%%%%%%%%%%%%%%%%%%%%%%%%%%%%%%
%%%%%%%%%%%%%%%%%%%%%%%%%%%%%%%%%%%%%%%%%%%%%%%%%%%%%%%%%%%%%%%%%%%%%%%%%%

\vspace{-0.3cm}
\section{Conclusions}
We have presented an application of the reweighting method to the case of the spatial periodicity of fermionic fields. We have studied 
the approach at tree level and by mean of numerical simulations. For the latter we have looked at the plaquette and at the pion
dispersion relation in small and large volumes. In both cases we have found the effects to be at the sub-percent level for values of $\theta$ up to $\pi/2$.

\end{document}